# Exact prediction of S&P 500 returns

Ivan O. Kitov, Oleg I. Kitov

**Introduction**

Let's assume that stock exchange represents an instrument for estimation of future states of real economies at various time horizons. Then, such aggregated indices as S&P 500 represent an estimate of the evolution of an economy as a whole. Accuracy of this estimate depends on the understanding of the causes of economic growth and the uncertainty associate with the measurements of real economic growth itself. When the future trajectory of real and nominal economic growth (both aggregated and disaggregated) is exactly known one can also predict stock market indices. Therefore, the problem of stock market prediction is the problem of the prediction of economic growth. Hence, accurate prediction of economic growth resolves two problems – stock exchange obtains a precise tool for estimation of future stock prices and the uncertainty of GDP measurements is also reduced. The discrepancy between measured GDP and that obtained from stock market indices should converge or, equivalently, macroeconomic state of any developed economy should be exactly described by aggregated stock market indices. The term "exactly" implies that increasing accuracy of the prediction of GDP unambiguously leads to vanishing discrepancy between predicted and observed (aggregated) stock market indices.

There is a link between the change rate of population of specific age and real economic growth as found by Kitov (2006a), Kitov, Kitov and Dolinskaya (2007). Real GDP in the USA has three principal components of growth. First of these components is associated with the extensive growth of working age population and thus workforce. This component of real economic growth is positive in the USA and reaches almost 1% per year. Therefore, when comparing the USA to other developed countries with low growth rate of population one should reduce USA figures by one percentage point per year. Effectively, this would mean comparison of real GDP per capita values, *GDPpc*. This variable defines second principal source of real economic growth. Kitov (2006b) found that developed economies (in the long-run) are characterized by constant annual increment of real GDP per capita - corresponding time series in these countries have no trend over the last 55 years of measurements (since 1950). Levels of these annual increments are very close in leading economies and define similar asymptotic behavior of real growth rate - inverse proportionality to the attained level of real GDP per capita. Economic trend is represented in the following form: *dln(GDPpc)=A/GDPpc*, where *A* is country specific constant.

Third source of growth is responsible for short-term fluctuations and is related to the change rate of single year of age population. This specific age in the USA is nine years as well as in the UK. Other European countries and Japan are characterized by specific age of eighteen years. The influence of the third source was tested using the examples of the USA, the UK, Japan, France and Austria. All these countries provide confidence in the validity of our approach.

For the USA, during the period between 1960 and 2005 real GDP growth is well described as a function of the three principal variables - working age population growth, constant increment of GDP per capita, and growth rate of nine-year-olds.

Working age population, WAP, and economic trend are slowly evolving variables. Between 1985 and 2005, the mean growth rate of WAP in the USA was 1.19% with the largest value of 1.5% in 2000 and the smallest value of 0.8% in 1989. Both extremes might be biased by revisions after censuses in 1990 and 2000. Economic trend in the USA, as defined by Kitov (2006b) and Kitov, Kitov, and Dolinskaya (2007), falls from 1.7 % in 1985 to 1.2% in 2005. In first approximation, we neglect small-amplitude changes in both variables and treat them as constants. One can extend our analysis and include actual behavior of these two variables. Despite slightly increased accuracy such an approach leads to a less parsimonious model.

Having tight relations between stock market indices and economic growth and between economic growth and specific age population one can test potential effects of the specific age population change on stock markets. The assumption behind such a link is very simple - the population change induces positive or negative economic growth, which, in turn, is reflected in stock prices. All other financial, economic, demographic and social factors are neglected! One can expect positive influence of the increasing specific age population on stock market indices. Moreover, one can predict stock market behavior long before any major change actually happens because population of the same year of birth does not change much over time. For example, one can accurately predict the number of 9-year-olds five years ahead using the current number of 4-year-olds. In other words, it is possible to predict large changes in stock market several years before they actually happen.

The remainder of the paper is organized as follows. Section 1 presents the prediction of the S&P 500 returns for the period between 1985 and 2003 as a function of measured and projected number of 9-year-olds. In Section 2, we test observed and predicted returns for cointegration and provide some estimates of prediction errors. Sections 3 and 4 predict the S&P 500 returns after 2003 using population estimates and real GDP growth rates, respectively. Section 5 concludes.

**1. Prediction of the S&P 500 returns between 1985 and 2003**

Before modeling S&P 500 (SP500) returns we would like to present some important features of population age distribution. The US Census Bureau (CB) has been providing monthly estimates of single year of age population since March 1990. Before 1990, only quarterly and annual estimates were available. The methodology of estimation, as described by the CB (2006b), includes monthly statistics of net births and deaths. Basic period for these estimates is one quarter, however. (Migration processes are described only at annual basis.) The absence of accurate monthly readings for the defining parameters of the population estimates potentially leads to a higher uncertainty and high order of autocorrelation due to smoothing or balancing of true changes through quarters and ages. Moreover, decennial censuses, which provide population counts at the ends of ten year intervals, are usually used to revise the whole evolution of population age pyramid across calendar years and ages. The latter revision additionally redistributes the counts in five- to ten-year-wide age groups in a way to reach an overall balance between adjacent ages. This procedure adds to the deterioration of the consistency of the estimates compared to true distribution.

Our aim is to describe SP500 returns using monthly readings of 9-year-old population in the USA. Due to the aforementioned problems with the accuracy of these particular monthly population estimates some smoothing is necessary in order to obtain a higher reliability. There are several representations tested for a better prediction of SP500 returns. One can smooth population readings using adjacent months or the same months of adjacent years. The former approach allows reducing large fluctuations induced by death statistics for 9-year-olds. The latter approach is inspirited by the methodology of the US Census Bureau which revises monthly and quarterly population estimates in specific age groups. It is also worth noting that these revisions depend on the error of closure which is different in various age groups in absolute and relative terms (US CB, 2006a). So, the best representation of monthly estimates may differ between age groups. One should be very careful in merging data from different age groups when interpolating the monthly estimates.

There is no doubt that high-frequency fluctuations of stock market prices are driven by a multitude of factors including news issued by political and financial authorities, changes in weather conditions, and individual actions of stock market participants. These short-term fluctuations have negligible effect on the long-term path of market indices, however. The latter is more likely to be governed by macroeconomic variables. Since the change in some specific age population is considered as a macroeconomic variable defining the evolution of real GDP per capita in developed countries one should model relatively long-term features of the indices but also take all the advantages of the high-frequency data set.

In this study, SP500 returns are represented by a sum of monthly returns during previous twelve months. Obviously, natural step in this (SP500 returns) time series is one month. This approach allows obtaining a smoother curve than that provided by annual returns with one month step, i.e. by the ratio of a given index level to that 12 month before. Figure 1 demonstrates the difference between these two definitions. Actually, we model modified annual SP500 returns with one month resolution.

Annual SP500 returns are modeled using monthly growth rates of smoothed 9-year-old population estimates. For example, the SP500 return for June 1995, i.e. of the sum of monthly returns between July 1994 and June 1995, is proportional to the ratio of monthly population estimates for May and June 1995 modified according to the procedure accepted in the study. Trail and error method indicated that the best monthly estimates of the number of 9-year-olds for our purposes are obtained when the estimates made for the same month but for five adjacent years of age are averaged. For example, the number of 9-year-old persons for June 1995 is estimated as the mean value of 7-, 8-, 9-, 10-, and 11-year-olds in June 1995. The intuition behind this representation is linked to the balancing of the number of single year of age population in five to ten year-wide age groups carried out by the US Census Bureau. Effectively, these revisions result in the interdependence of monthly estimates for adjacent ages. The approach used in the study potentially recovers a part of true monthly values, but this problem has to be investigated in more detail.

The approximation of the number of 9-year-olds, $N9(t)$, is justified also by the excellent prediction of the SP500 returns, $R_p(t)$, for the period where monthly estimates are available, except the years after 2003. Linear model linking SP500 return and the change rate of the number of 9-year-olds has the following form:

$$R_p(t) = AdlnN9(t) + B, \qquad (1)$$

where coefficients *A* and *B* are determined empirically because they are dependent on the differences in corresponding change rates and the smoothing and weighting procedures applied by the US Census Bureau to the monthly population estimates. Notice, that coefficient *B* approximates inputs of working age population growth and economic trend, as discussed in Introduction.

There are two time series for the number of 9-year-olds between 1990 and 2000: *post*censal and *inter*censal. The former is obtained by "inflation-deflation" component method using contemporary estimates of total deaths and net migration including net movement of the US Armed Forces overseas. The start point for corresponding single year of age time series is the values counted in the 1990 Census. The *inter*censal time series is estimated using the population counts in the 1990 and 2000 censuses as the start and end points. The single year of age *inter*censal time series are obtained from corresponding postcensal series by a proportional redistribution of the errors of closure over the ten years between the censuses on the daily basis. Since the errors of closure are age dependent, adjacent time series may converge or diverge by several per cent. For example, the *post*censal estimate of the population under five years of age for April 2000 was underestimated by about 1% relative to the counted value, the population between 5 and 13 years of age was underestimated by 3.5%, and that between 14 and 17 years of age was underestimated by 2.4%. Additional disturbance to the monthly estimates of the age pyramid is introduced by the adjustment of the sum of the single year estimates to the total population obtained by an independent procedure. After April 2000, only *post*censal population estimates are available. There are several vintages of these estimates are available for previous years, however, which use most recent information on past estimates of death rate and migration. So, no *post*censal estimate is final and further revisions are probable for any monthly estimate.

When a single year of age population is used for the prediction of SP500 returns, the difference between the *post-* and *inter*censal populations is expressed only in synchronized and proportional change in level. The difference in the change rate due to the error of closure is evenly distributed over the months and practically not visible for monthly estimates. There is a several percent difference for the cumulative curves as dictated by the error of closure for the 9-year-olds.

As discussed above, the (*post*censal) monthly estimates of 9-year-olds in our study are obtained using also the numbers of 7-, 8-, 10-, and 11-year-olds. All these ages are inside the US Census Bureau specified age group between 5 and 13 years. Using the monthly estimates of *N9(t)* one can predict SP500 returns according to (1). Figure 2 displays observed and predicted time series for SP500 returns for the period between 1990 and 2000. The latter is obtained by varying coefficient *A* and *B* to minimize the RMS difference between two time series. No formal minimization procedure was used, however, and the best manually obtained coefficients *A*=170 and *B*=-0.04 provide RMS of 0.082 with mean value of -0.003 for the difference. Visually, the predicted and measured curves are similar. In the long run, high-frequency fluctuations in both series are cumulated to zero, as Figure 3 demonstrates. Average SP500 return (according to the definition accepted in our study) for the same period is 0.16 with standard deviation of 0.10. The predicted time series is characterized by average value of 0.158 and standard deviation of 0.091.

Figure 4 represents a predicted curve obtained using the *inter*censal estimate of 9-year-olds between 1991 and 2003. Three years between 2000 and 2003 are obviously *post*censal estimates, but they are potentially of high accuracy due to their closeness to the single year of age counts of the 2000 Census. The best fit coefficients *A*=165 and *B*=-0.055 provide RMS difference of 0.085. The mean measured SP500 return for the period between 1990 and 2003 is 0.15 with standard deviation of 0.10.

The *post-* and *inter*censal estimates of the number of 9-year old provide a consistent description of the SP500 returns between 1991 and 2003. There is a good opportunity, however, to obtain a relatively accurate prediction of these returns at time horizons somewhere between one and nine years using population estimates for younger ages as a proxy to the number of 9-year-olds. For example, the number of 7-year-olds provides a good approximation for monthly increment, and thus change rate, of 9-year-olds, as Figure 5 illustrates. The procedure we have developed includes averaging of five consecutive single year population estimates between 7 and 11 years of age for obtaining the estimates of the number of 9-year-olds. Hence, the estimate of 7-year-olds includes the ages from 5 to 9, but two years before.

Figure 6 displays the measured and predicted (using 7-year-olds) curves for SP500 returns. The latter is a forecast at a two-year horizon using the estimates for younger ages. RMS difference between these two time series for the period from 1992 to 2003 is only 0.088 with the best-fit coefficients *A*=165 and *B*=-0.061. Therefore, at a two-year horizon one can obtain a prediction with the uncertainty of 0.088 for the SP500 (yearly) return. When using current population estimates, it is possible to extend the forecasting horizon to nine years, with slightly degrading accuracy however. Population projections allow obtaining even longer predictions of SP500 returns.

Standard econometric and financial approach denies predictability of stock markets considering the evolution of stock markets as a random innovation or random walk process with some trend component. At a two-year horizon, the SP500 return time series is characterized by standard deviation of 0.18, i.e. twice as large as the one obtained in the predicted time series. Effectively, these values demonstrate the predictability of the SP500 returns what is of great importance for any stock market participant. Moreover, significant improvement in the prediction is available through the improvement in the accuracy of population estimates.

Having monthly population estimates after April 1990 one can extrapolate the description of the SP500 returns in the past using older population. The intuition behind this approach is the same as for the prediction of the future returns using younger ages - the monthly increments change slowly for the same year of birth population.

First, the procedure of the averaging of five adjacent ages for representing monthly estimates has been tested for older ages. Unfortunately, it gave poorer results compared to those for the period between 1990 and 2003. The deterioration of the results might be associated with the differences in the revision procedures applied by the US CB to the population estimates in the older age groups between 14 and 17 years and between 18 and 24 years. For the extrapolation of the 9-year-olds time series by four and more years in the past the younger one or both of the age groups have to be used in the five-year wide averaging interval. Therefore, an alternative approach was used, which is based on averaging of adjacent months for the same age. When 12 successive moths are used, this approach is identical to the obtaining of the cumulative SP500 return for the previous 12 months. For shorter averaging windows, however, the result is very similar to that using 12 months, as Figure 7 demonstrates. For the purpose of the prediction of SP500 returns before 1991, we used a four-month wide averaging window and the 17-year of age population estimate. Figure 8 depicts the measured and predicted SP500 return between 1984 and 1995. The latter is shifted by eight years back relative to its natural position and obtained with the best-fit coefficients *A*=35 and *B*=0.089, which are apparently different from those obtained by averaging of monthly estimates for five consecutive years. Despite high-frequency fluctuations, the predicted curve repeats the most prominent features of the measured one. Notice an almost precise prediction of time and amplitude of the stock market crash in 1987! The difference between readings of the measured and predicted curves is presented in Figure 9 and characterized by a finite variance and zero mean. RMS value of the difference for the period from 1985 to 1991 is only 0.096, which is much better than that obtained using a naïve (random walk) prediction at an eight-year horizon.

Monthly estimates of SP500 returns provide a dynamic view, which is characterized by high-frequency fluctuations not related to the long-term equilibrium link between the stock market index and the number of 9-year-olds. Figures 3 and 9 demonstrate that the difference between the measured and observed SP500 returns has a zero mean and no linear trend. Therefore, one can expect that in the long run the differences are canceled out and corresponding cumulative curves have a strong tendency to converge and demonstrate the unbiased long-term relationship between the measured and observed SP500 returns. Figure 10 displays these cumulative curves, which are actually characterized by a few small-amplitude deviations, which are compensated at short time intervals. The most prominent features such as periods of near-zero and negative returns are well described, however. The periods of rapid growth are also well predicted. All these features are modeled using only one parameter - the number of 9-year-olds. This is an ultimately parsimonious model, which also provides accurate forecasts at various time horizons. The period of such an excellent description finished in April 2003, however.

## 2. Cointegration test

Despite the similarity between the measured, $R_m(t)$, and predicted, $R_p(t)$, SP500 returns, both dynamic and cumulative, formal econometric tests may additionally validate the link between the stock market index and the evolution of specific age population. In this Section, we test the existence of a long-term equilibrium (cointegrating) relation between the measured and predicted SP500 returns during the period between 1985 and 2003. As shown in Section 1, there is no reliable link between these variables before 1985 and after 2003, i.e. one can not extend the above obtained empirical relationships with available monthly population estimates beyond those dates. The period after 2003 will be analyzed and modeled in the next two Sections.

According to Granger and Newbold (1967), the technique of linear regression for obtaining statistical estimates and inferences related to time series is applicable only to stationary time series. Two or more nonstationary series can be regressed only in the case when there exists a cointegrating relation between them (Hendry and Juselius, 2001). Therefore, the first step in the econometric studies of time dependent data sets is currently consists is estimation of the order of integration of involved series. Unit root tests applied to original series and their first and higher order differences are a useful tool to determine the order of integration.

There are four time series to be tested for unit roots - the measured and predicted SP500 returns and their first differences. Standard econometric package Stata9 provides a number of appropriate procedures implemented in interactive form. The Augmented Dickey-Fuller (ADF) and the modified DF t-test using a generalized least-squares regression (DF-GLS) are used in this study. These tests provide adequate results for the available series consisting of 207 monthly readings – SP500 returns for previous 12 months between 1985 and 2003.

Corresponding results of unit root tests for these four series are listed in Tables 1 and 2. Both original series are characterized by the presence of unit roots - the test values are significantly larger than 5% critical values. Both first differences have no unit roots and thus are stationary. In the ADF tests, trend specification is constant and the maximum lag order is 3. In the DF-GLS tests, the maximum lag is 4 and the same trend specification is used.

The presence of unit roots in the original series and their absence in the first differences evidences that the former series are integrated of order 1. This fact implies that a cointegration analysis has to be carried out before any linear regression because such regression potentially is a spurious one.

The assumption that the measured and predicted (i.e. the change rate of 9-year-olds) returns are two cointegrated non-stationary time series is equivalent to the assumption that their difference, $\varepsilon(t) = R_m(t) - R_p(t)$, is a stationary or I(0) process. Therefore, it is natural to test the difference for unit root. If $\varepsilon(t)$ is a non-stationary variable having a unit root, the null of the existence of a cointegrating relation can be rejected. Such a test is associated with the Engle-Granger approach (1987), which requires $R_m(t)$ to be regressed on $R_p(t)$, as the first step. It is worth noting, however, that the predicted variable is obtained by a procedure similar to that of linear regression and provides the best visual fit between corresponding curves.

The Engle-Granger approach is most reliable and effective when one of the two involved variables is weakly exogenous, i.e. is driven by some forces not associated with the second variable. This is the case for the SP500 returns and the number of 9-year-olds. The latter variable is hardly to be driven by the former one.

The results of the ADF and DF-GLS tests listed in Table 3 indicate the absence of unit roots in the difference between the measured and predicted series. Since the predicted series are constructed in the assumption of a zero average difference, trend specification in the tests is *none*. The maximum lag order in the tests is 3. The test gives strong evidences in favor of the existence of a cointegrating relation between the measured and predicted time series. Thus, from econometric point of view, it is difficult to deny that the number of 9-year-olds is potentially the only defining factor behind the observed long-term behavior of SP500.

The Johansen (1988) approach is based on the maximum likelihood estimation procedure and tests for the number of cointegrating relations in the vector-autoregressive representation. The Johansen approach allows simultaneous testing for the existence of cointegrating relations and determining their number (rank). For two variables, only one cointegrating relation is possible. When cointegration rank is 0, any linear combination of the two variables is a non-stationary process. When rank is 2, both variables have to be stationary. When the Johansen test results in rank 1, a cointegrating relation between involved variables does exist.

In the Johansen approach, one has first to analyze specific properties of the underlying VAR model for the two variables. Table 4 lists selection statistics for the pre-estimated maximum lag order in the VAR. Standard trace statistics is extended by several information criteria: the final prediction error, FPE, the Akaike information criterion, AIC, the Schwarz Bayesian information criterion - SBIC, and the Hannan and Quinn information criterion, HQIC. All tests and information criteria indicate the maximum pre-estimated lag order 3 for VARs and VECMs. Therefore, maximum lag order 3 was used in the Johansen tests along with constant as trend specification.

Table 5 represents results of the Johansen tests – in both cases cointegrating rank is 1, i.e. there exists a long-term equilibrium relation between the measured and predicted (i.e. the number of 9-year-olds) SP500 returns. We do not test for causality direction between the variables because the only possible way of influence, if it exists, is absolutely obvious.

Now we are sure that the measured and predicted time series are cointegrated. Therefore, the estimates of the goodness-of-fit and RMSE in various statistical representations have to be valid and provide important information on the accuracy of corresponding measurements and the relation itself. VAR representation is characterized by $R^2$=0.89 and RMSE=0.047 (with mean annual return of 0.18 for the same period) due to strong noise suppression. In practice, AR is a version of a weighted moving average, which optimizes noise suppression throughout the whole series. Simple linear regression provides a lower $R^2=0.66$ and larger RMSE = 0.07.

### 3. Prediction of S&P 500 returns after April 2003

We have to admit that the period after April 2003 is not described well when monthly population estimates are used. The measured SP500 return started to grow in April 2003. By February 2004 it increased by 0.61 – from -0.28 to +0.33. There is no sign of such an increase in the 9-year-old postcensal estimates, however. Younger ages, some of them explicitly counted during the 2000 Census, also do not demonstrate any significant steps in 2003.

Let's analyze what might be the cause for the failure to describe the SP500 return after April 2003. Was there a structural break in the stock market behavior or corresponding population estimate is wrong? Both assumptions allow any evolution of the SP500 returns after 2003. The former assumption has quantitative consequences. It presumes the existence of a new relationship between SP500 returns and the number of 9-year-olds after 2003. If the population estimates are inherently inaccurate no modeling is possible.

Therefore, we have tried to model the current development of the SP500 returns using the same specific-age population approach from some point after 2003. The short period of large increase in SP500 returns after 2003 was excluded from the prediction and was approximated by a step in January 2004. Effectively, the period between April 2003 and July 2004 is not modeled.

In any quantitative prediction, dynamic range of the change of linked variables plays a crucial role. When two variables are not changing with time one can not reveal any link between them. When both variables are changing in a wide range one can obtain a reliable estimate of the link between them. The period after July 2004 is characterized by relatively small oscillations in SP500 returns. This fact creates significant problems for calibration of any relationship between the SP500 returns and the number of 9-year-oldsin the USA. Nevertheless, some relative changes are available. In that sense any linear relationship between SP500 returns and the change rate is valid and scalable. The latter property can be used when large changes in SP500 returns will occur.

The first estimate for the period after 2004 was made in January 2007. It provided a crude prediction of some important events in 2007 and 2009. These two events were strong growth of the SP500 returns with sudden and very sharp drop in 2007 and 2009. We developed the following model for the period after 2005: $R_p=30dln(N9)-0.1$, where the number of 9-year-olds is approximated by the number of 3-year-olds with a 6-year shift, i.e. one can predict at a 6-year horizon. Figure 12 displays some details of the current period and gives monthly predictions of the SP500 returns after 2007. As mentioned above, there is a severe drop in 2007 followed by an 11-month period of growth with annual returns around 0.05. Monthly population estimates indicate that this fall should be in October-November 2007. A strong and short rally in 2008 will raise the SP500 returns at the level of 0.2 to 0.25 and will end in a catastrophic fall down to 0.1. All in all, the next 2 to 3 years have to validate our model for the period after 2004.

Figure 13 demonstrates the difference between the measured and predicted SP500 returns for the period between 1985 and 2010. This difference has no linear trend and is characterized by normal distribution. Figure 14 provides a long-term view on the evolution of the SP500 returns. After the period of a decrease between 2001 and 2003, the SP500 return is growing at a constant pace in the long run. Crudely, this growth continues the trend observed between 1985 and 1996. Thus, one can consider the period between 1996 and 2003 as a deviation from the overall trend consisting of a segment of abnormal growth (1996-2001) and fast decrease (2001-2003). This evidences in favor of fundamental factors defining the long-term trend of the stock market growth.

4. **S&P 500 returns and real GDP**

In Section 3, we have demonstrated that a structural break possibly occurred in 2003. This break might be associated with two different mechanisms. First, the break is induced by some real economic processes, i.e. may reflect the change in the inherent link between true values of the studied variables – the number of 9-year-olds and SP500 returns. This new link is likely to be linear, as is the link observed before 2003. In such a case, the relationship obtained in Section 3 for the period after 2004 is valid before some new structural break will occur.

Second mechanism is related to some changes in population measuring procedure. In this case, the structural break is artificial and the relationship for the period before 2003 can be easily transported in a scaled version to the period after 2003. Is there a possibility to distinguish between these two mechanisms for the change in 2003?

Originally, the link between real GDP growth rate and the change rate of the number of 9-year-olds was found (Kitov, 2006a). Corresponding relationship should work in both directions, i.e. one can estimate the growth rate of real GDP from population measurements, and the number of 9-year-olds from real GDP measurements. In Section 3, we failed to predict the step-like increase in the SP500 returns in 2003 and 2004 using population estimates, as they are presented by the US Census Bureau. There is a possibility that these estimates are not accurate enough because of the increasing distance from the 2000 census and are obtained by some modified methodology.

So, in relationship (1), one can replace *N9(t)* with *GDPpc(t)*, taking into account that second term in the relationship between real GDP per capita and population is constant. Figure 15 displays the observed SP500 returns and those obtained using real GDP, as presented by the US Bureau of Economic Analysis (www.bea.gov). As before, the observed returns are 12-month cumulative values. The predicted returns are obtained from the relationship

$$R_p(t) = 10.0*dln(GDPpc(t)) - 0.25,$$

where *GDPpc(t))* is represented by the mean (annualized) growth rate during two previous quarters. Only quarterly readings of real GDP are available.

The period after 1996 is well predicted including the sharp increase in 2003. Therefore, it is reasonable to assume that the 9-year-old population was not well estimated by the US Census Bureau after 2003. This conclusion is supported by the cointegration test conducted for real GDP per capita and the charge rate of the number of 9-year-olds (Kitov, Kitov, Dolinskaya, 2007), which proves the existence of a long-term equilibrium linear relation between these two variables since the early 1960s. As a result, one can use either *N9(t)* or *GDPpc(t)* for the modeling of the SP500 returns, where one of them is more appropriate. Obviously, the *GDPpc(t)* is consistent with the SP500 returns after 2003. The year after second quarter of 2006 is not well predicted, however.

There is a general concern related to the accuracy of the most recent measurements of population and real GDP. In Figure 15, the predicted curve dropped to -0.075 in the third quarter of 2006 in accordance with low growth rate of real GDP. There was no drop in the SP500 returns during the same period. A possible reason for the discrepancy is the underestimation of real GDP growth in the past two years. New revisions of these readings should be dramatically large (+1 to +1.5 percentage points), if the link between the real GDP and SP500 holds.

On the other hand, the number of 9-year-olds also demonstrates the possibility of a large decrease in the SP500 returns starting late in 2007. Real GDP should follow the same path as the long-term relation between these variables prescribes. Therefore, the drop in the SP500 returns is very likely as both population and real GDP measurements indicate.

The success in modeling of the SP500 returns using *GDPpc* indicates that the population estimates after 2003 are obtained according to a somewhat revised procedure. Therefore, the prediction in Section 3 is likely to be valid. The years between 2007 and 2010 should confirm or reject this statement.

**5. Conclusion**

One can derive many theoretical and practical conclusions from the above analysis. We consider the following findings as principal:

- Annual SP500 returns can be exactly predicted at time horizons from 1 month to 9 years.
- The accuracy of this prediction depends only on the accuracy of the estimates of the number of 9-year-olds.
- The period between 2007 and 2010 is very important for the validation of our model since contains severe changes in the number of 9-year-olds.

**References**

Engle, R., Granger, C. (1987). Cointegration and error correction: representation, estimation, and testing. Journal of Econometrics, 55, 251-276

Granger, C., Newbold, P. (1967). Spurious regression in econometrics. Journal of Econometrics, 2, 111-120

Hendry, D., Juselius, K. (2001). Explaining Cointegration Analysis: Part II. Energy Journal, 22, 75-120

Johansen, S. (1988). Statistical analysis of cointegrating vectors. Journal of Economic Dynamics and Control, 12, 231-254

Kitov, I.O. (2006a). GDP growth rate and population, Working Papers 42, ECINEQ, Society for the Study of Economic Inequality. www.ecineq.org/milano/WP/ECINEQ2006-42.pdf

Kitov, I.O., (2006b). Real GDP per capita in developed countries, MPRA Paper 2738, University Library of Munich, Germany.
http://mpra.ub.uni-muenchen.de/2738/01/MPRA_paper_2738.pdf

Kitov, I.O., Kitov, O.I., Dolinskaya, S., (2007). Modelling real GDP per capita in the USA: cointegration test, MPRA Paper 2739, University Library of Munich, Germany.
http://mpra.ub.uni-muenchen.de/2739/01/MPRA_paper_2739.pdf

U.S. Census Bureau. (2006a). Methodology. United States Population Estimates by Age, Sex, Race, and Hispanic Origin Method: July 1, 2006. Retrieved February 26, 2007 from http://www.census.gov/popest/topics/methodology/2006_nat_meth.html

U.S. Census Bureau. (2006b). National intercensal estimates (1990-2000). Retrieved February 26, 2007 from http://www.census.gov/popest/archives/methodology/ intercensal_nat_meth.html

**Tables**

Table 1. Results of unit root tests for original time series – measured and predicted SP500 returns. Both series are characterized by the presence of unit roots.

| Test   | Lag | predicted | measured | 1% critical |
|--------|-----|-----------|----------|-------------|
| ADF    | 0   | -2.65     | -2.07    | -3.47       |
|        | 1   | -2.58     | -1.63    | -3.47       |
| DF-GLS | 1   | -2.69     | -2.47    | -3.48       |
|        | 2   | -2.34     | -1.99    | -3.48       |

Table 2. Results of unit root tests for the first differences of the original time series – measured and predicted SP500 returns. Both series are I(0).

| Test | Lag | predicted | measured | 1% critical |
|---|---|---|---|---|
| ADF | 0 | -15.6* | -16.6* | -3.47 |
|  | 1 | -12.0* | -9.2* | -3.47 |
|  | 2 | -9.2* | -7.8* | -3.47 |
|  | 3 | -8.7* | -7.7* | -3.47 |
| DF-GLS | 1 | -10.9* | -9.1* | -3.48 |
|  | 2 | -8.1* | -7.6* | -3.48 |
|  | 3 | -7.4* | -7.5* | -3.48 |
|  | 4 | -6.2* | -7.7* | -3.48 |

Table 3. Results of unit root tests of the differences of the predicted and measured time series. There is no unit root in the difference.

| Test | Lag | difference | 1% critical |
|---|---|---|---|
| ADF | 0 | -7.6* | -3.47 |
|  | 1 | -6.8* | -3.47 |
|  | 2 | -6.4* | -3.47 |
|  | 3 | -6.8* | -3.48 |
| DF-GLS | 1 | -6.7* | -3.48 |
|  | 2 | -6.3* | -3.48 |
|  | 3 | -6.7* | -3.48 |

Table 4. . Pre-estimation lag order selection statistics. All tests and information criteria indicate the maximum lag order 3 as an optimal one for VARs and VECMs.

| Lag | LR | FPE | AIC | HQIC | SBIC |
|---|---|---|---|---|---|
| 0 |  | 0.0065 | -2.19 | -2.18 | -2.16 |
| 1 | 211 | 0.0023 | -3.22 | -3.20 | -3.17 |
| 2 | 1.39 | 0.0023 | -3.21 | -3.19 | -3.15 |
| 3 | 9.5* | 0.0022* | -3.254* | -3.22* | -3.17* |
| 4 | 1.7 | 0.0022 | -3.253 | -3.21 | -3.15 |

Table 5. Johansen test for cointegration rank for the measure and predicted time series. Maximum lag order is 3.

| Trend specification | Rank | Eigenvalue | SBIC | HQIC | Trace statistics | 5% critical value |
|---|---|---|---|---|---|---|
| none | 1 | 0.196 | -5.79* | -5.93* | 2.93* | 3.84 |
| rconstant | 1 | 0.196 | -5.65* | -5.92* | 3.24* | 9.25 |
| constant | 1 | 0.196 | -5.74* | -5.90* | 2.67* | 3.76 |

**Figures**

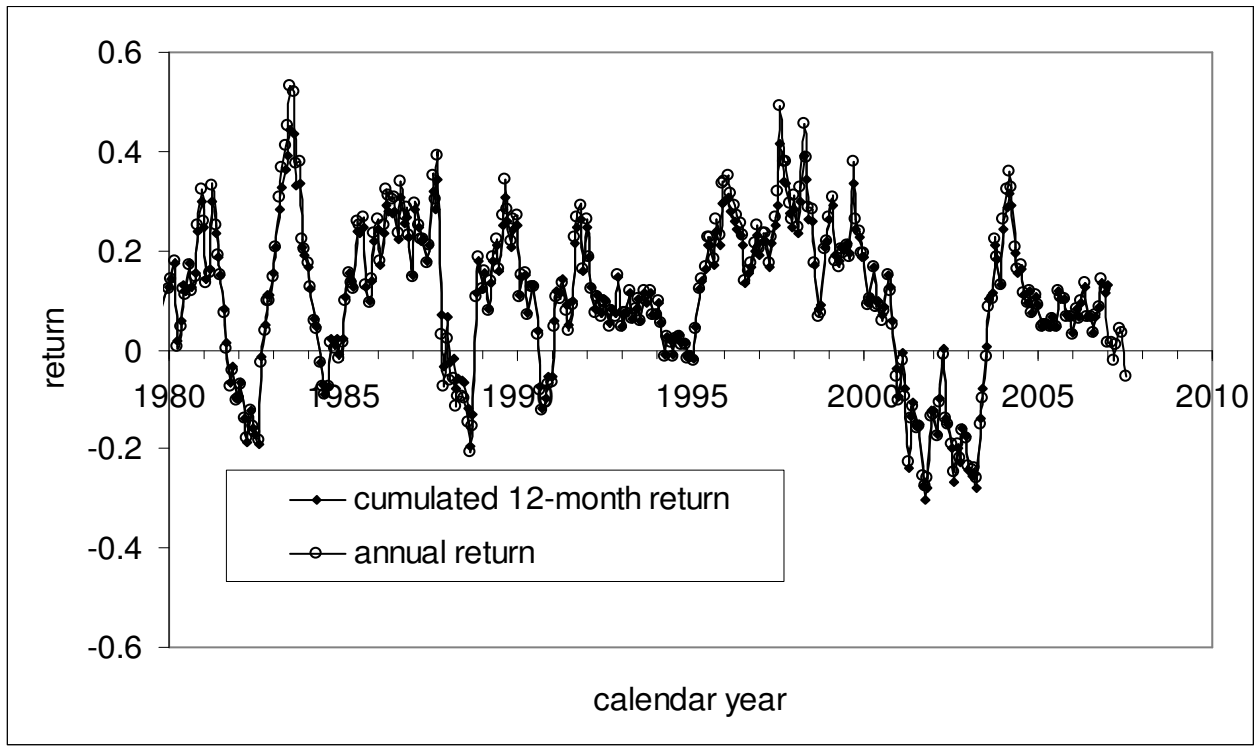

Figure 1. Comparison of annual SP500 return and that cumulated during the previous twelve months as a sum of monthly returns. The annual curve is of a slightly higher volatility.

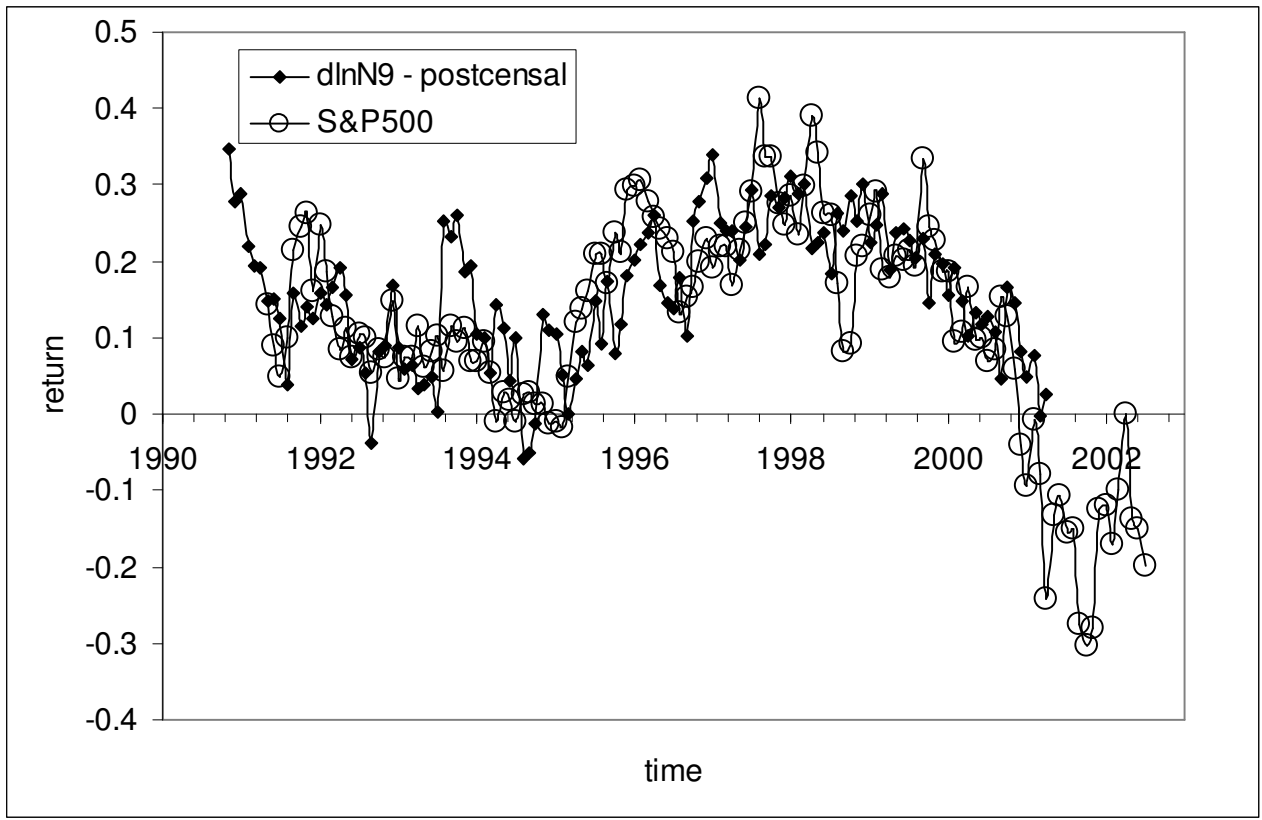

Figure 2. Comparison of observed and predicted SP500 returns between 1990 and 2000. The latter is obtained using the *post*censal estimate of the 9-year-olds. RMS difference between the curves for the period between 1991 and 2001 is 0.082 with mean value only -0.003. Corresponding coefficients are: *A*=170 and *B*=-0.004.

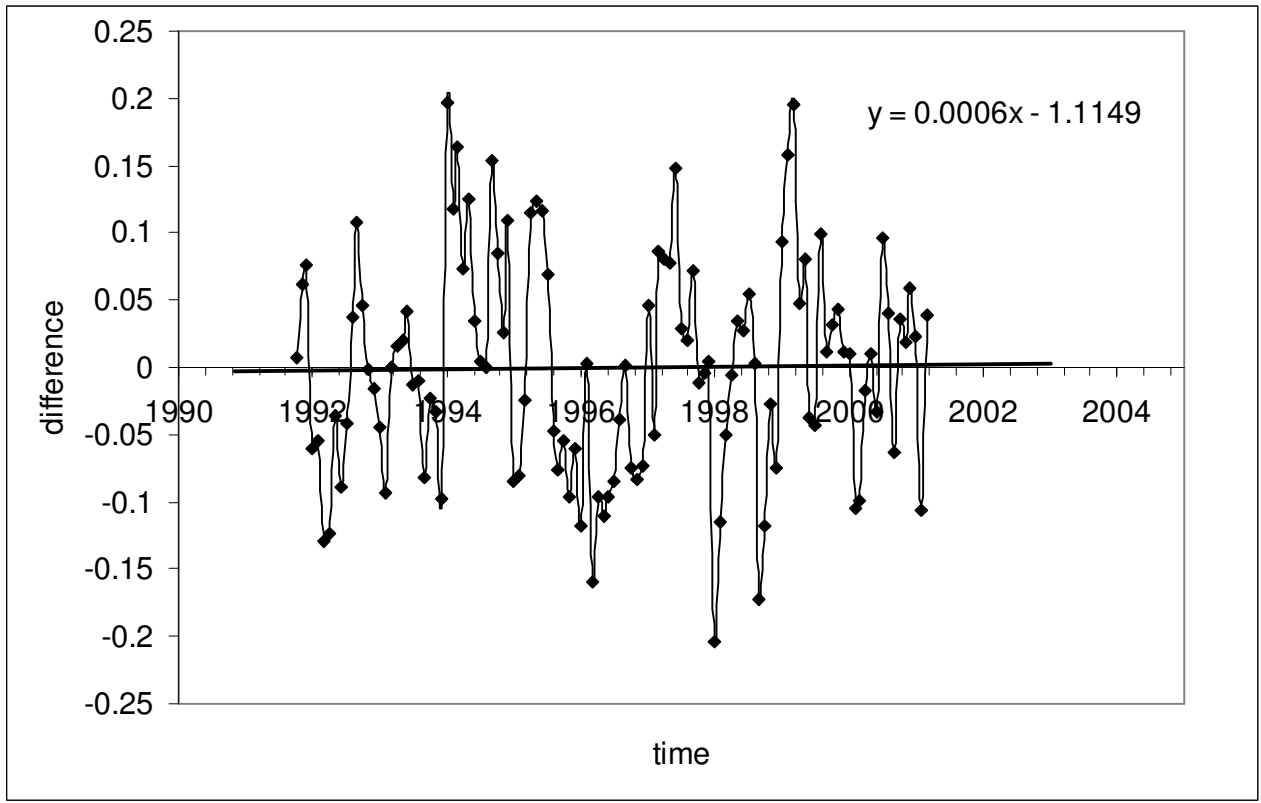

Figure 3. The difference between the measured and observed SP500 presented in Figure 2. In the long run, the difference practically has no linear trend.

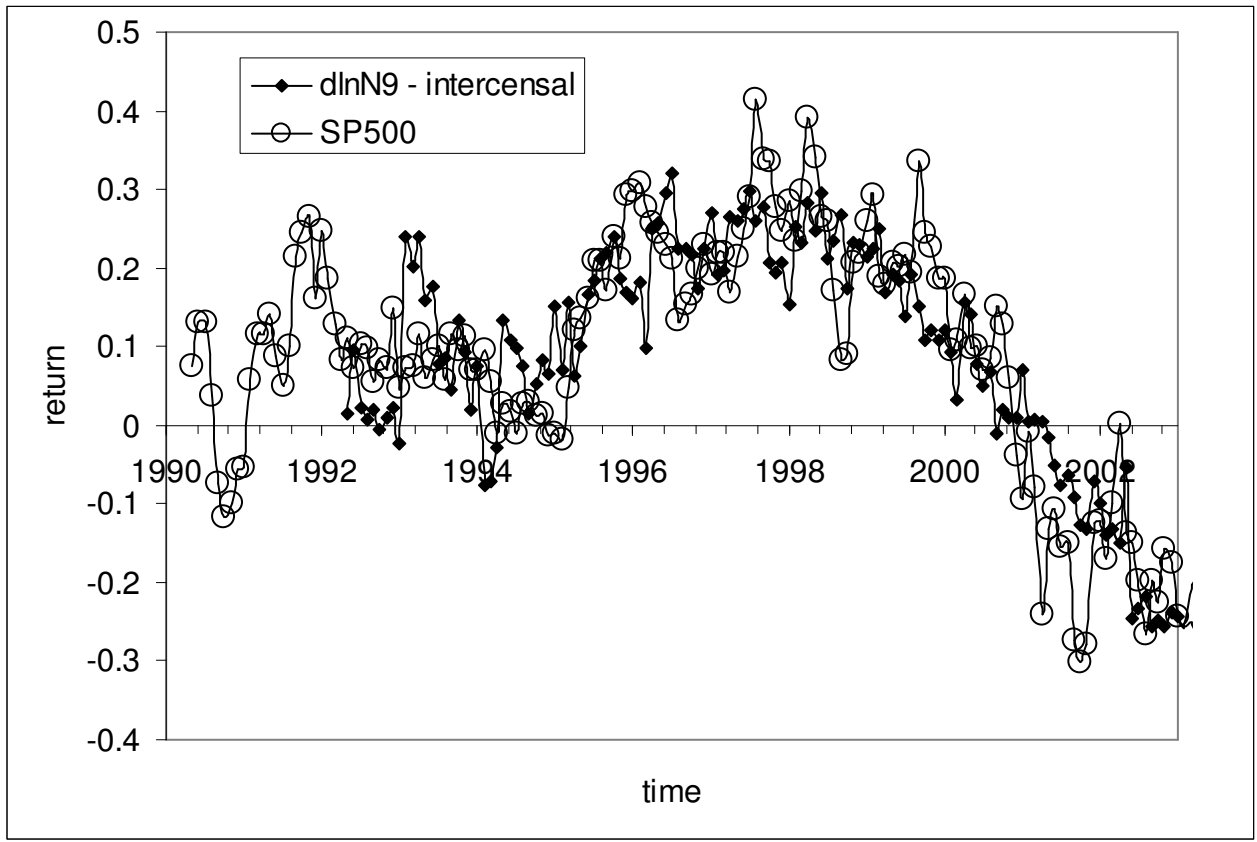

Figure 4. Comparison of the observed and predicted SP500 returns between 1992 and 2003. The latter is obtained using the *inter*censal estimate of the 9-year-olds with *A*=165 and *B*=-0.055.

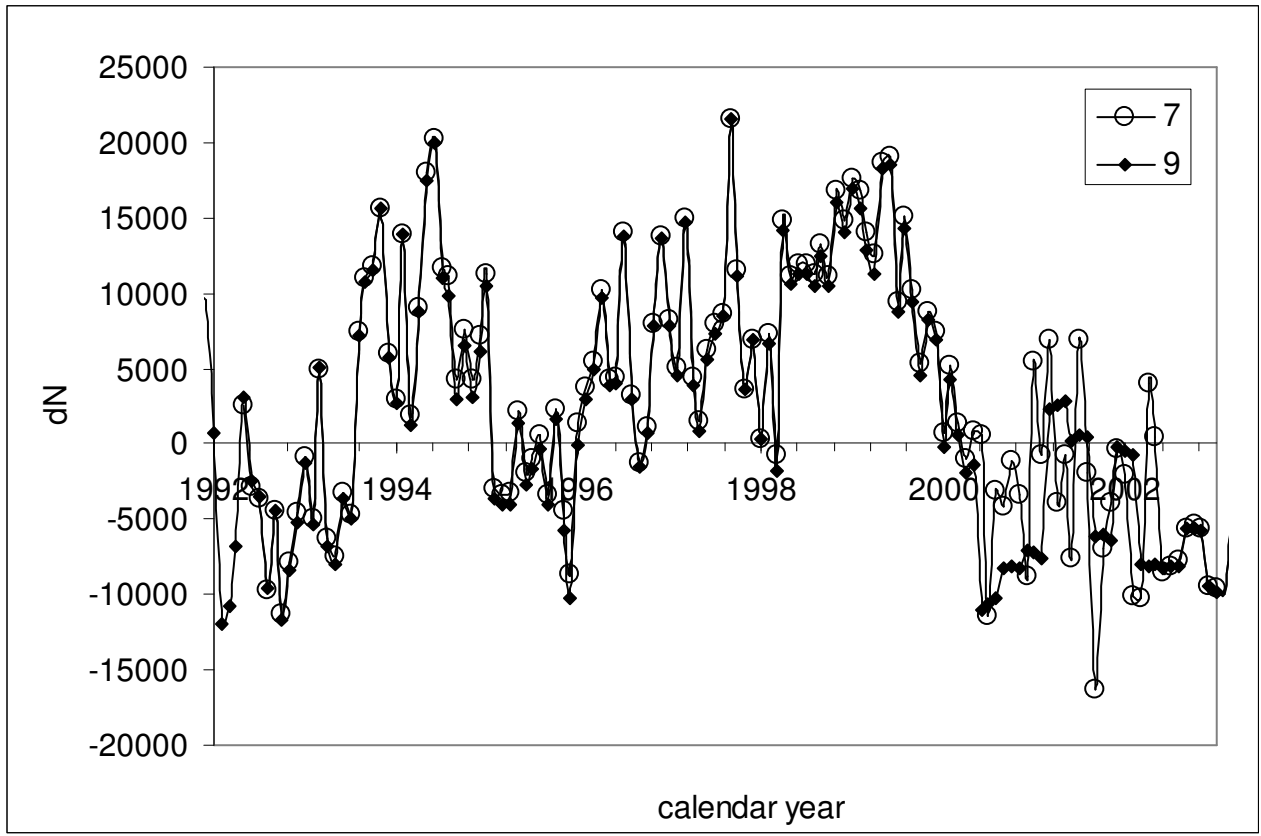

Figure 5. Comparison of monthly increments in the number of 9-year-olds and the number of 7-year-olds shifted two years ahead.

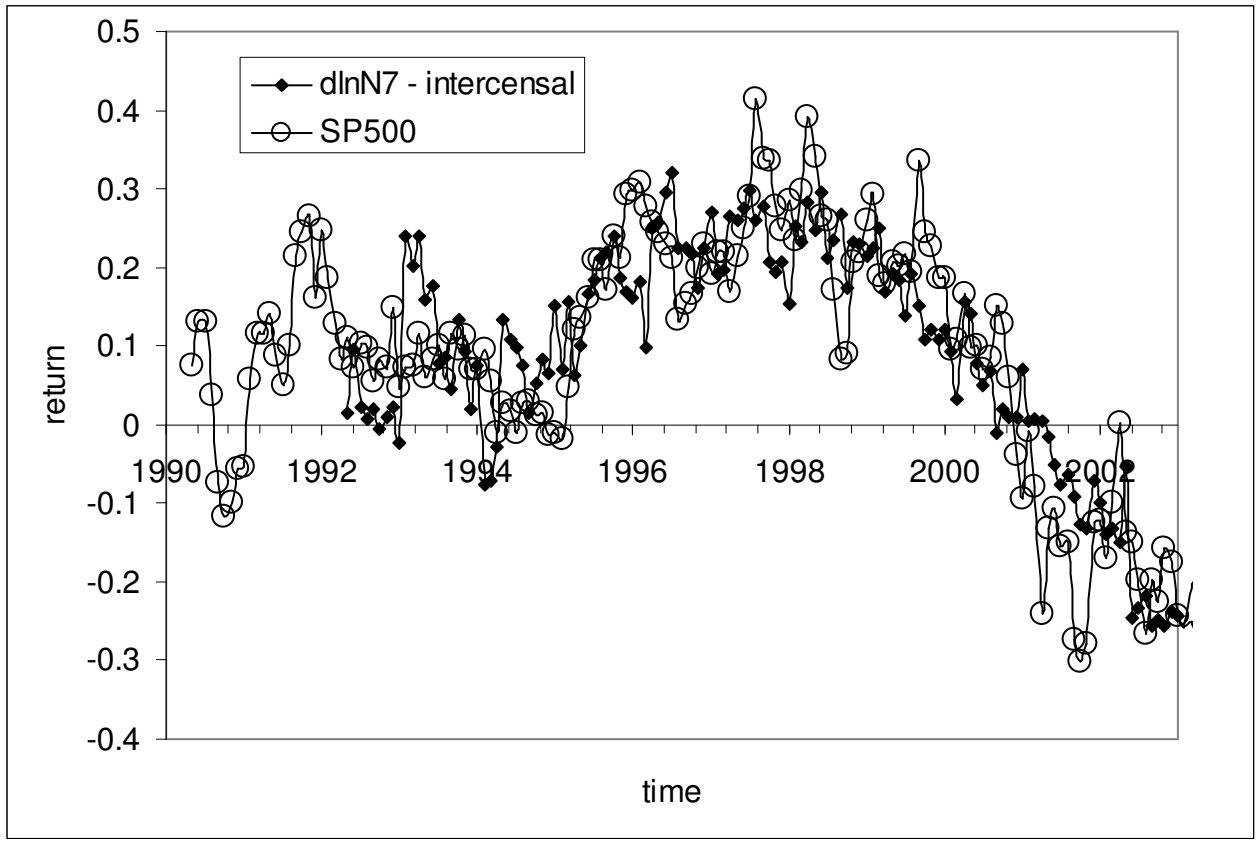

Figure 6. Comparison of the observed and predicted SP500 returns between 1992 and 2003. The latter is obtained using the *inter*censal estimate of the 7-year-olds shifted two years ahead with *A*=165 and *B*=-0.061.

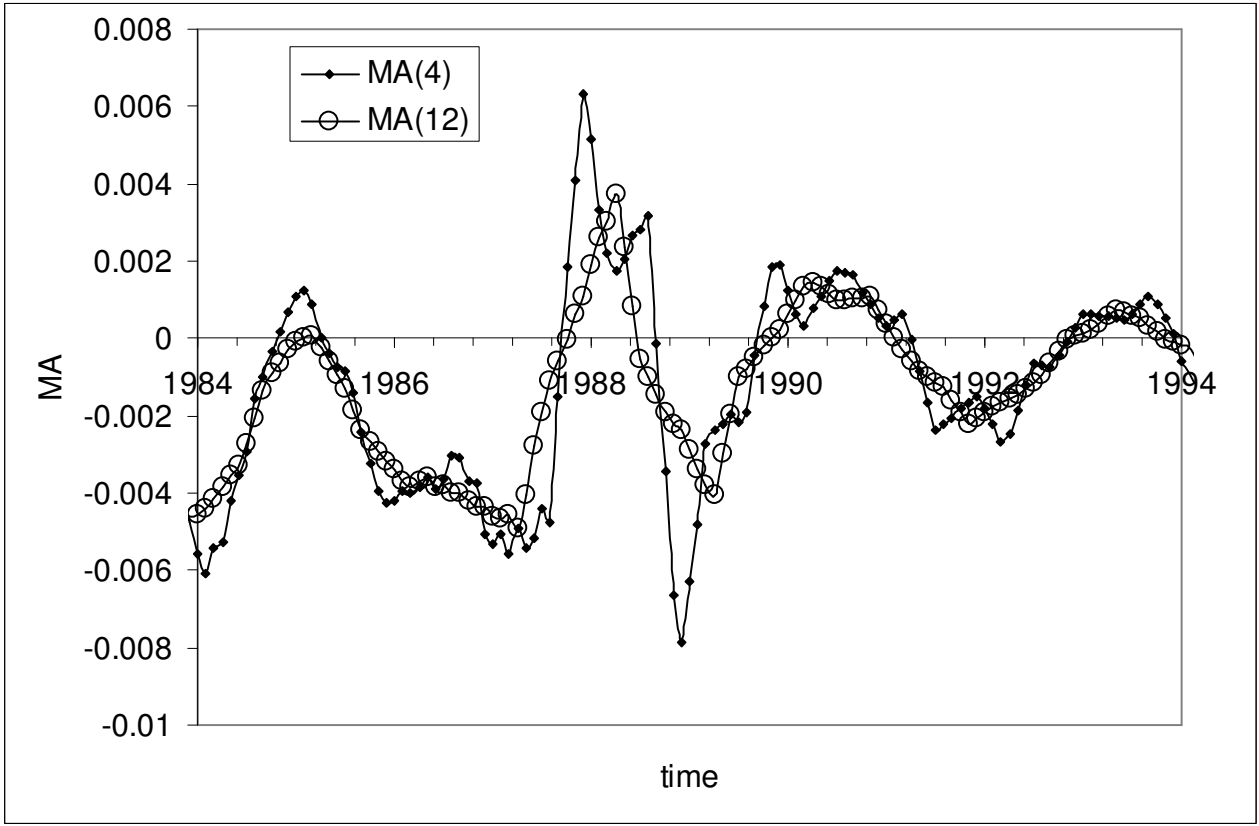

Figure 7. Comparison of 4- and 12-month moving averages of the number of 17-year-olds.

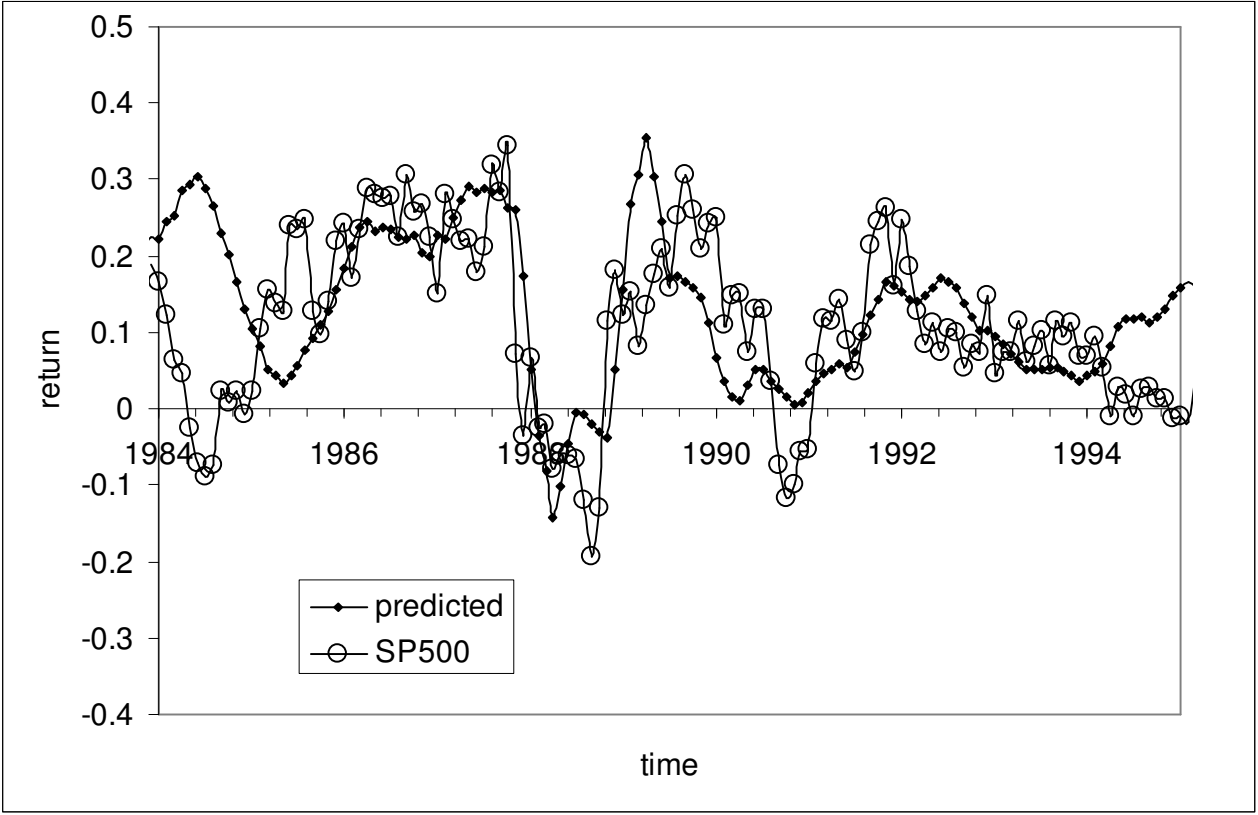

Figure 8. Comparison of the observed and predicted SP500 returns between 1984 and 1995. The latter is obtained using the *inter*censal estimate of 17-year-olds.

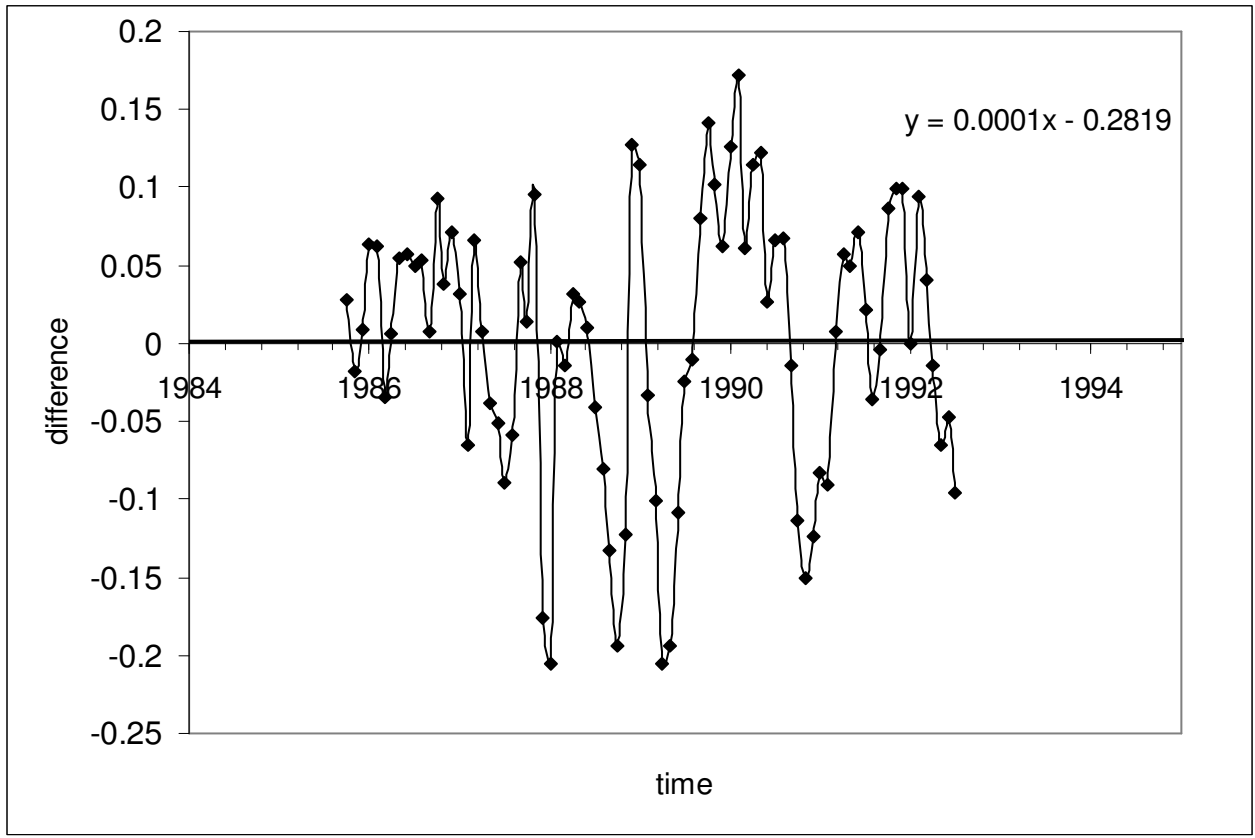

Figure 9. The difference between the measured and predicted SP500 returns for the period between July 1985 and July 1992.

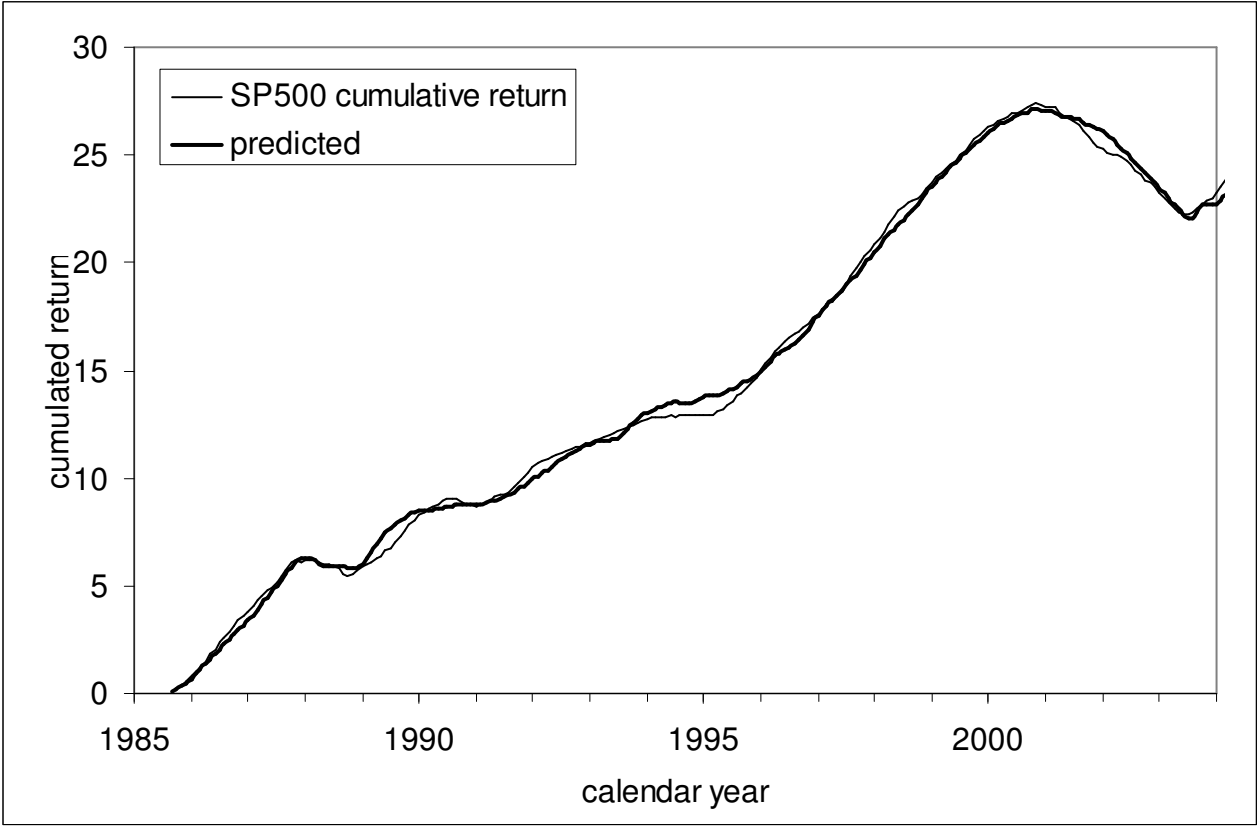

Figure 10. Comparison of the measured and predicted cumulative SP500 returns presented in Figures 2 and 8.

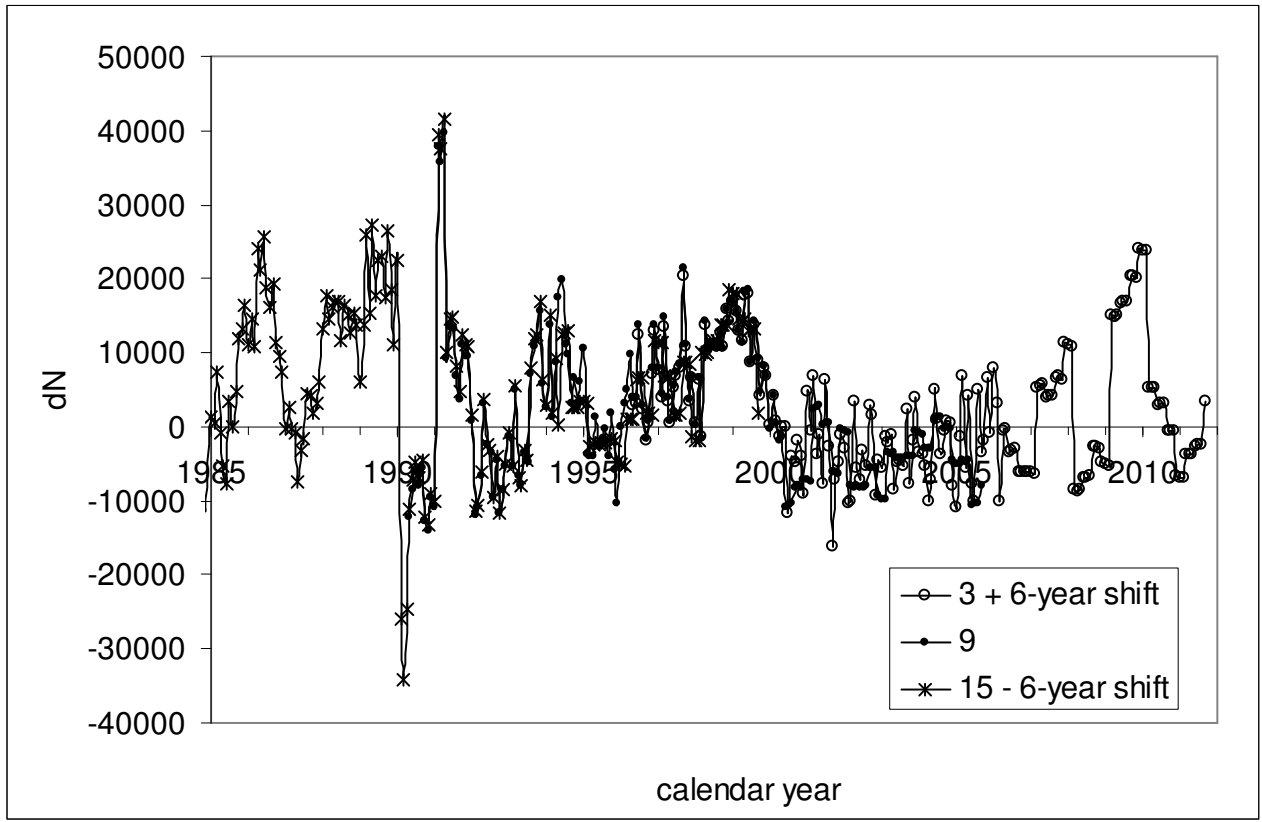

Figure 11. The change rate of the number of 9-year-olds between 1985 and 2010. Before 1990 the number of 9-year-olds is represented by the number of 15-year-olds shifted by 6 years back, and after 2005 by the number of 3-year-olds shifted six years ahead.

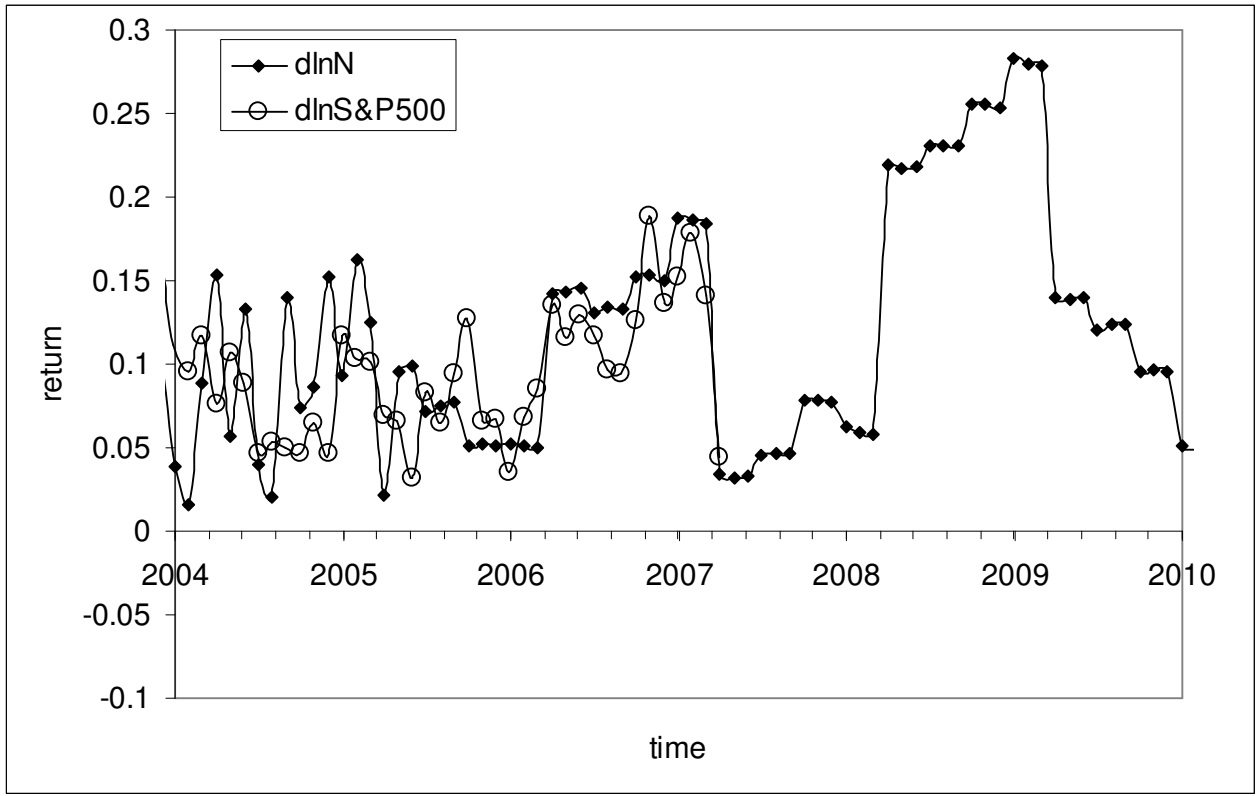

Figure 12. Comparison of the observed and predicted SP500 returns between 2004 and 2010. The predicted curve is obtained from the 3-year-old estimates in Figure 11 without any time shift.. Notice perfect timing of the fall in November 2007.

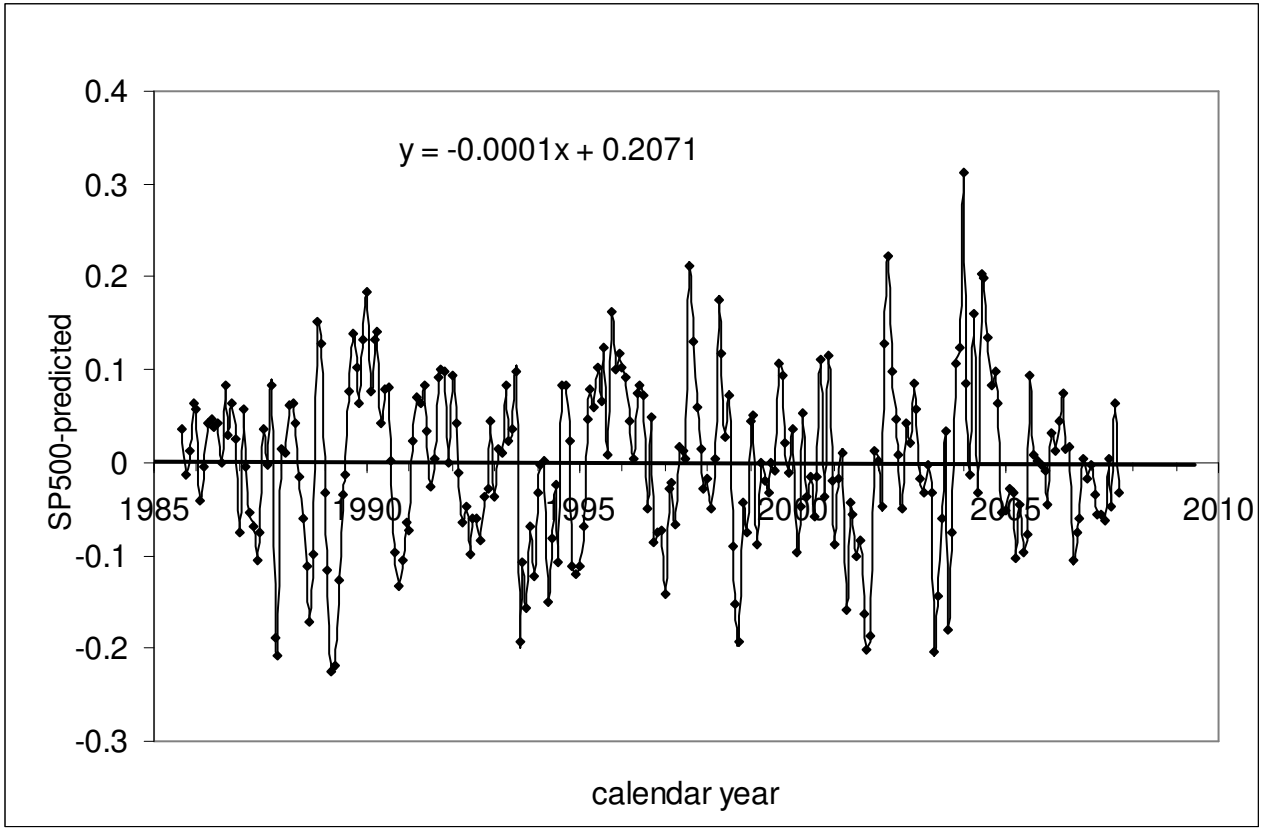

Figure 13. The difference between the measured and predicted SP500 return for the period between 1985 and 2007.

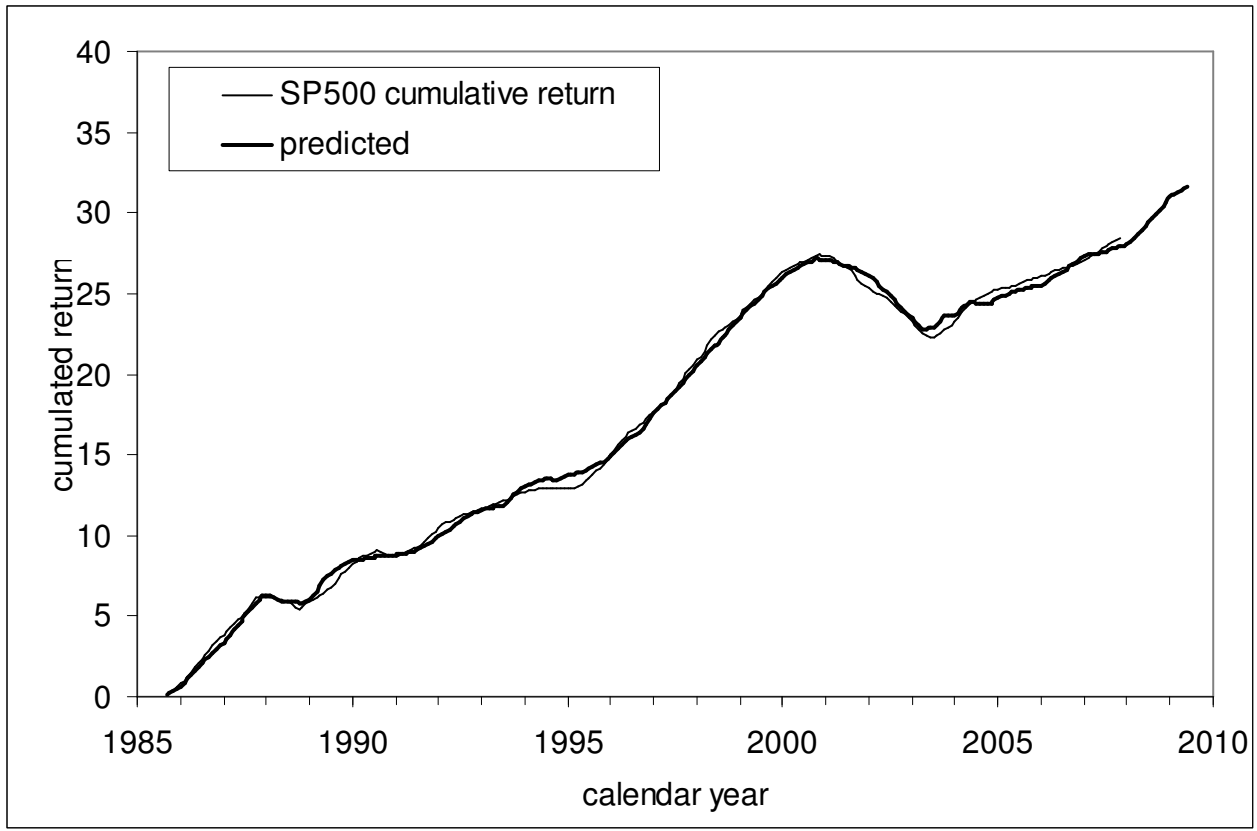

Figure 14. Comparison of the measured and predicted cumulative SP500 return presented in Figure 11.

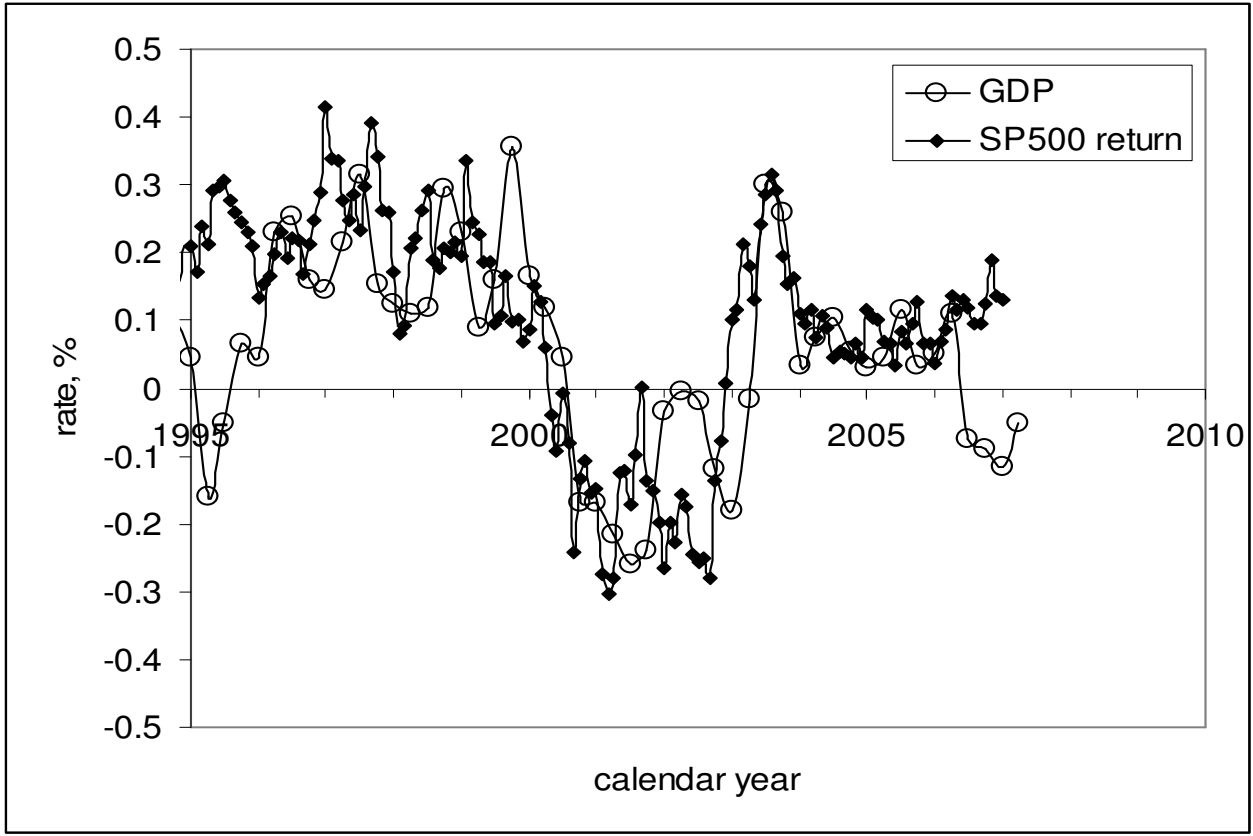

Figure 15. The observed and predicted SP500 returns. The latter are obtained using quarterly readings of the growth rate of real GDP.